\documentclass[a4paper]{article}

\usepackage[english]{babel} \usepackage{hyperref} \usepackage{float}
\usepackage[utf8]{inputenc} \usepackage{amsmath} \usepackage{graphicx}
\usepackage[colorinlistoftodos]{todonotes} \usepackage{tikz}
\usepackage{pdfpages} \usepackage{listings}
\usepackage{afterpage}
\usepackage{lscape}
\usepackage{caption}
\usepackage[a4paper,margin=1in]{geometry}
\usepackage{authblk} 
\usepackage{hyperref} 
\usepackage{titlesec} 
\usepackage{lmodern}
\usepackage{tabularx} 
\usepackage{array}
\usetikzlibrary{arrows,positioning,shapes.geometric}

\titleformat{\section}{\large\bfseries}{\thesection}{1em}{}

\title{Design and Implementation of an IoT Cluster with Raspberry Pi Powered by Solar Energy: A Theoretical Approach}

\author[1]{Noel Portillo\thanks{\href{https://orcid.org/0009-0009-8817-3187}{ORCID: 0009-0009-8817-3187}}}
\affil[1]{\textit{Axxis Technologies}, Houston, TX, USA \\
\href{mailto:noel@axxistechnologies.com}{noel@axxistechnologies.com}}
\affil[2]{\textit{Independent Researcher}, Ahuachapán, El Salvador \\
\href{mailto:noel.portillo@ieee.com}{noel.portillo@ieee.com}}

\date{September 25, 2024}

\begin{document} \maketitle

\section{Abstract}

This document presents the design and implementation of a low-power IoT server cluster, based on Raspberry Pi 3 Model B and powered by solar energy. The proposed architecture integrates Kubernetes (K3s) and Docker, providing an efficient, scalable, and high-performance computing environment.

The cluster is designed to optimize energy consumption, leveraging a 200W solar panel system and a 100Ah lithium-ion battery to support continuous operation under favorable environmental conditions.

Performance analysis was conducted based on theoretical inferences and data obtained from external sources, evaluating resource allocation, power consumption, and service availability. These analyses provide theoretical estimates of the system’s operational feasibility under different scenarios.

The results suggest that this system can serve as a viable and sustainable alternative for edge computing applications and cloud services, reducing dependence on traditional data centers.

In addition to its positive impact on environmental sustainability by significantly reducing the carbon footprint, this solution also addresses economic concerns, as conventional data centers consume enormous amounts of energy, leading to increased demand on the power grid and higher operational costs.
 
\section{Introduction}

The Internet of Things (IoT) is transforming the way computing systems operate, enabling automation and real-time processing in industrial, residential, and enterprise environments. Its applications extend to various technological fields such as Machine Learning, Artificial Intelligence (AI), and microservices, both in local infrastructures and the cloud. An example of this trend is the solution proposed by Souza et al., 2024\cite{ref1}, which implements a distributed system based on a Raspberry Pi cluster for monitoring a photovoltaic plant.

Traditional data centers consume vast amounts of energy, much of which still comes from fossil fuels, leading to high operational costs and a significant environmental impact. In this context, this study proposes the implementation of an IoT server cluster powered by solar energy, aiming to reduce energy consumption, optimize operational costs, and minimize the carbon footprint.

Although Raspberry Pi-based clusters have been previously explored in educational and research settings, few implementations have focused on energy efficiency and the use of renewable energy sources. This paper presents an integrated solution, combining open-source software, containerized services, and an optimized energy management strategy, demonstrating the feasibility of this approach for edge computing applications and distributed services.

\section{Problem Statement}
In 2019, the world faced one of the worst public health crises in modern history: the COVID-19 pandemic. This event tested traditional business and communication models, highlighting the urgent need to adopt new technologies. The COVID-19 pandemic significantly impacted small businesses, particularly those dependent on in-person customer interactions. As a result, many companies were forced to rapidly adopt digital technologies to maintain operations.

The pandemic drove a rapid digitalization process in small businesses, increasing the adoption of technology from 30\% to 55\% in the United States, according to reports from small business owners (Petrovskaya \& Kiseleva, 2024)\cite{ref2}. However, accessibility to technology remains a significant barrier, particularly in developing regions or among small family-owned businesses. Essential digital services such as email hosting, e-commerce platforms, and inventory management software introduce ongoing operational costs, which many businesses struggle to afford.

Additionally, the implementation of traditional on-premises servers requires an upfront investment ranging from \$500 to \$1,500 USD, depending on hardware specifications and software licensing. On the other hand, cloud-based alternatives, such as email services averaging \$6.00 USD per user per month and e-commerce solutions like Shopify (approximately \$49.00 USD per month), increase operational expenses, adding additional financial constraints for small businesses.\cite{ref4}.

Given these challenges, an alternative solution is the implementation of low-cost, open-source computing platforms powered by renewable energy. This research explores the feasibility of deploying an IoT cluster using Raspberry Pi devices with solar power, providing an energy-efficient, scalable, and economically viable computing infrastructure.

In contrast, a viable alternative for small businesses is the adoption of open-source software solutions on low-cost hardware. For example, a 5th generation Raspberry Pi, priced at around \$120.00 USD, could support the deployment of these services without incurring expensive licensing fees. This approach not only democratizes access to technology but also significantly reduces operational costs, facilitating small businesses' digital transformation and allowing them to compete in the digital economy without relying on costly infrastructure.
\section{Objectives}

\begin{enumerate}
    \item Develop a theoretical framework for building a \textit{computational cluster prototype}, utilizing \textit{low-cost development boards} such as the \textit{Raspberry Pi}.
    
    \item Assess the technical feasibility and sustainability of implementing a \textit{computational cluster}, leveraging \textit{open-source software} and a \textit{solar-powered energy system}.
    
    \item Disseminate the obtained results to facilitate the analysis, optimization, and replication of the model, enabling its application by the \textit{scientific community} and researchers in the fields of \textit{distributed computing and renewable energy}.
\end{enumerate}

\section{Prototype Architecture}

The proposed system integrates various hardware and software components to ensure a scalable, efficient, and secure infrastructure.

\subsection{Hardware}

The cluster infrastructure consists of the following hardware components, selected for their \textit{energy efficiency} and \textit{compatibility with embedded environments}. Table~\ref{tab:hardware} summarizes the key hardware components used in the prototype.

\begin{table}[h]
    \centering
    \caption{Hardware Components of the Prototype}
    \label{tab:hardware}
    \begin{tabular}{|l|l|}
        \hline
        \textbf{Component} & \textbf{Specifications} \\ 
        \hline
        \multicolumn{2}{|c|}{\textbf{Cluster Nodes}} \\
        \hline
        5 × Raspberry Pi 3 Model B & Quad-core ARM Cortex-A53 @ 1.2 GHz, \\
                                   & 1 GB LPDDR2 SDRAM, Wi-Fi 802.11b/g/n/ac, \\
                                   & Bluetooth 4.2/BLE, Ethernet 10/100 BaseT, \\
                                   & 4 × USB 2.0, HDMI 1.4, Composite Video, \\
                                   & CSI for camera, DSI for touch display, \\
                                   & 3.5 mm audio jack, MicroSD card reader, \\
                                   & 40-pin GPIO, Dimensions: 88 × 56 × 19.5 mm, \\
                                   & Power consumption: 3.5 W per unit \\  
        \hline
        \multicolumn{2}{|c|}{\textbf{Networking and Connectivity}} \\
        \hline
        5 × Patch cord Cat 5e & 1 ft \\ 
        1 × Ethernet to USB 3.0 Adapter & Gigabit Ethernet 10/100/1000 Mbps \\ 
        1 × NETGEAR GS305 Switch & 5-port Gigabit Ethernet 10/100/1000 Mbps \\ 
        \hline
        \multicolumn{2}{|c|}{\textbf{Storage}} \\
        \hline
        1 × Storage shield for Raspberry Pi & MSATA SSD, 2 slots \\ 
        1 × MSATA SATA III SSD & 1 TB \\ 
        5 × MicroSD Cards & 32 GB each \\ 
        \hline
        \multicolumn{2}{|c|}{\textbf{Power and Energy}} \\
        \hline
        2 × Solar Panels & 100W each \\ 
        1 × Lithium Battery & 20Ah \\ 
        1 × MPPT Solar Charge Controller & - \\ 
        1 × Voltage Regulator & 12V to 5V, 8A \\ 
        \hline
        \multicolumn{2}{|c|}{\textbf{Other Components}} \\
        \hline
        5 × Micro USB Cables & - \\ 
        1 × Raspberry Pi Mounting Kit & - \\ 
        \hline
    \end{tabular}
\end{table}

\subsection{Hardware Configuration}

The cluster operates based on an \textit{Ethernet network}, where nodes communicate through a \textit{Gigabit network switch}. One of the nodes is designated as the \textit{Master Node}, responsible for \textit{managing, coordinating, and supervising resources} across the worker nodes in the cluster, dynamically allocating them as required by the user.  

This layer also incorporates \textit{external SSD storage}, allowing for \textit{expanded cluster storage capacity} and improved data handling performance.  

Although the prototype is based on the \textit{Raspberry Pi 3 Model B}, this choice was driven by \textit{availability and cost constraints} rather than scalability limitations. Currently, \textit{IoT boards with higher processing power and memory} are available and could be integrated into future iterations of the cluster. One such example is the \textit{Orange Pi}, an open hardware project similar to the Raspberry Pi, which offers more advanced configurations \cite{ref5}.  

Furthermore, since many of these board designs are \textit{publicly accessible}, the possibility of developing a \textit{custom hardware design} is not excluded, allowing for enhanced capabilities tailored to the specific needs of each application or project.  

\begin{figure}[H]
\centerline{\includegraphics[width=0.7\linewidth, keepaspectratio]{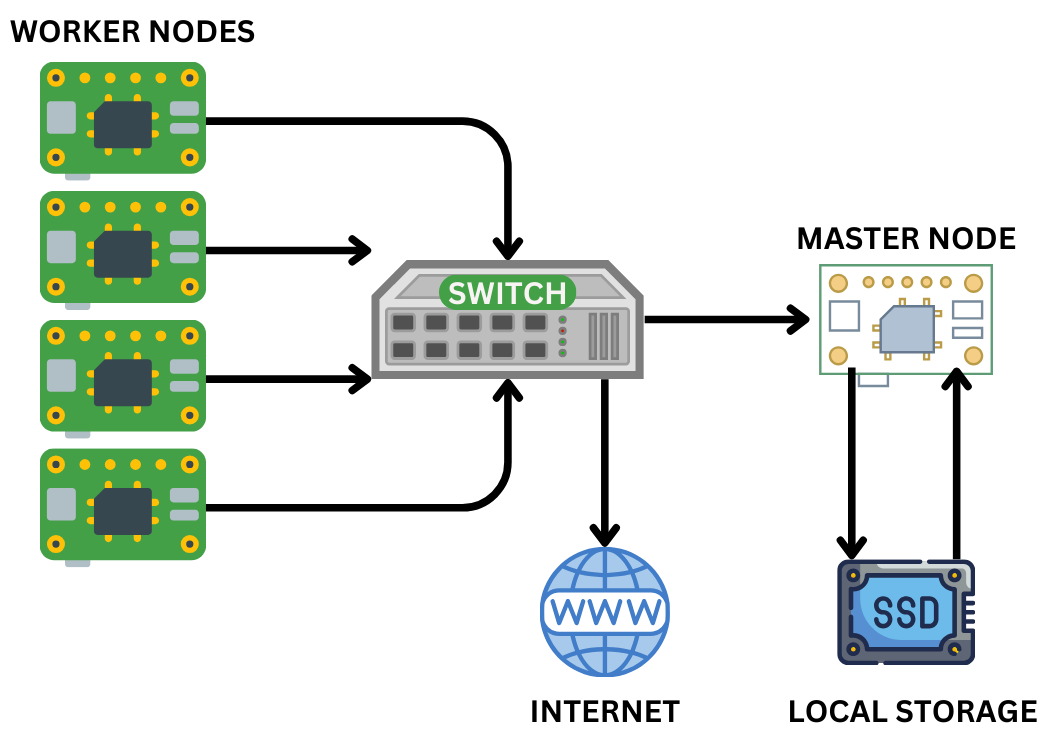}}
\captionof{figure}{Hardware Configuration of the Prototype}
\label{prototype_hardware_setup}
\end{figure}

\subsection{Software Configuration}

There are various alternatives for \textit{configuring the cluster}, depending on its intended use. For instance, if the cluster is designed for \textit{intensive mathematical processing}, a \textit{distributed computing} scheme can be implemented using an operating system optimized for ARM architectures, such as \textit{Raspberry Pi OS}. For container orchestration and load distribution, \textit{K3s}, a lightweight version of Kubernetes, is recommended, while \textit{MPI (Message Passing Interface)} is suitable for inter-node communication. In \textit{distributed computing}, libraries like \textit{NumPy + Dask} are particularly useful, while for \textit{Machine Learning}, \textit{TensorFlow} is an essential tool.  

However, in this solution, the primary objective is to provide \textit{web microservices}, such as \textit{email, marketing platforms, and e-commerce systems}. Therefore, the selected software configuration is as follows:  

\begin{table}[h]
    \centering
    \caption{Software Components of the Cluster}
    \label{tab:software}
    \begin{tabular}{|l|l|}
        \hline
        \textbf{Component} & \textbf{Software} \\ 
        \hline
        Operating System & Ubuntu Server 22.04 (ARM64) \cite{ref6} \\ 
        Container Orchestration & K3s \cite{ref7} \\ 
        Load Balancing & Traefik \cite{ref8} \\ 
        Web Server & Nginx \cite{ref9} \\ 
        Database Engine & MariaDB (MySQL derivative) \cite{ref10} \\ 
        Email Services & Postfix + Dovecot \cite{ref11, ref12} \\ 
        E-commerce & WooCommerce (WordPress-based) \cite{ref13} \\ 
        Marketing Platform & Mautic \cite{ref14} \\ 
        Monitoring & Prometheus + Grafana + Loki \cite{ref15, ref16, ref17} \\ 
        Security & Fail2Ban + UFW \cite{ref18, ref19} \\ 
        \hline
    \end{tabular}
\end{table}

It is important to highlight that \textit{all software solutions used in this project are open-source (GPL, GNU licensed)}, significantly reducing operational costs by \textit{eliminating the need for expensive proprietary licenses} in both configuration and production environments.

\begin{figure}[H]
\centerline{\includegraphics[width=0.5\linewidth, keepaspectratio]{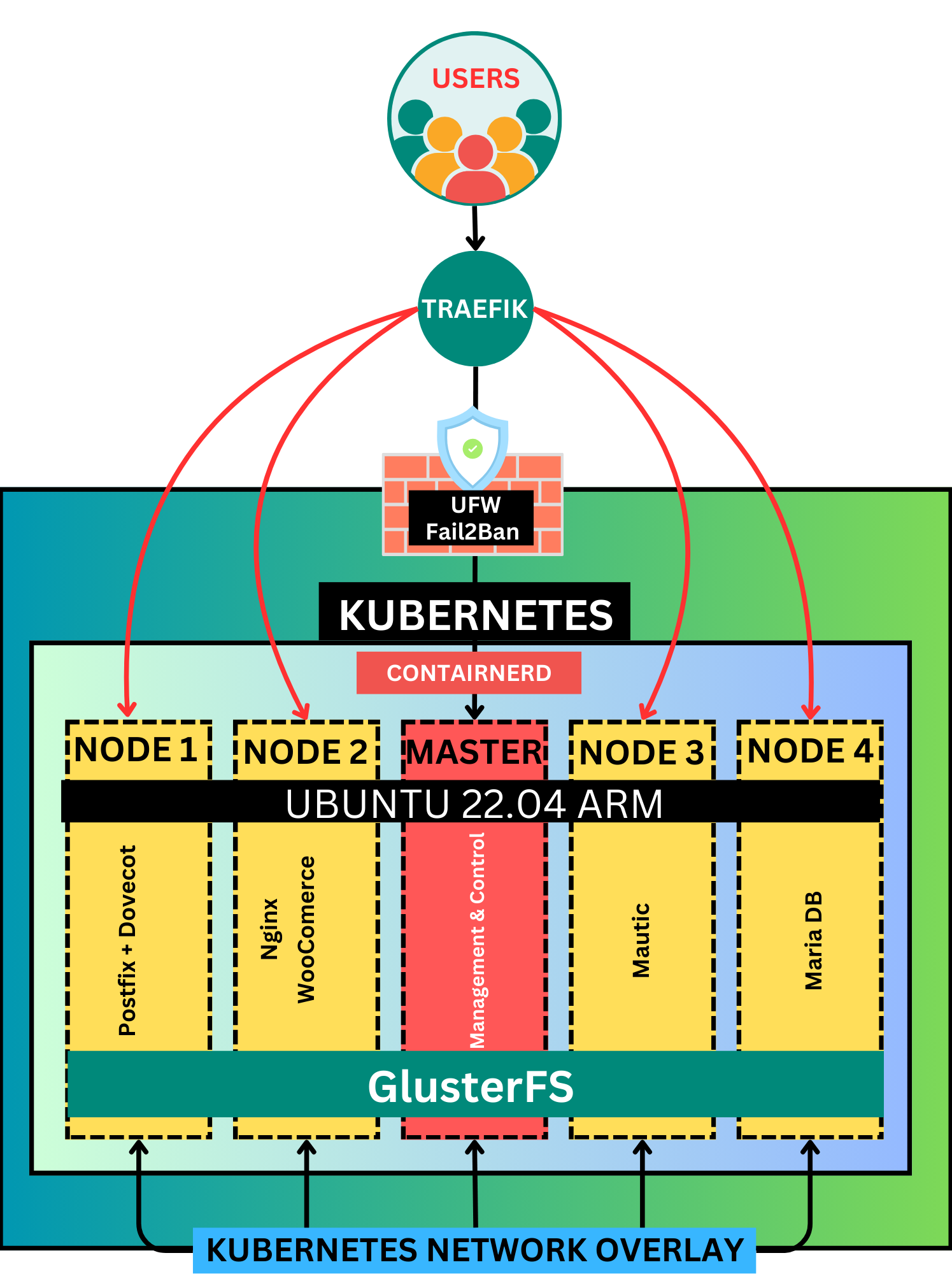}}
\captionof{figure}{Software Configuration of the Prototype}
\label{prototype_software_setup}
\end{figure}

\section{Energy}

One of the most innovative aspects of this project, in addition to the use of \textit{low-cost, energy-efficient IoT devices}, is the integration of \textit{renewable energy} as the primary power source.  

Currently, the world faces major challenges in \textit{energy sustainability and environmental impact reduction}. Companies providing \textit{technology services and solutions} require \textit{vast amounts of electricity} to operate their data centers, most of which is still derived from \textit{fossil fuels}. In addition to generating \textit{high carbon emissions}, these fuels have experienced \textit{rising costs due to geopolitical issues}, limiting their viability as a sustainable energy source.  

For this reason, this project prioritizes the \textit{use of solar energy and other renewable sources} to power the IoT cluster, aligning with the vision of a \textit{sustainable and accessible infrastructure}. This strategy not only reduces \textit{dependence on polluting energy sources} but also provides a \textit{viable solution for small business owners}, who, in addition to accessing \textit{advanced technological tools}, can lower their \textit{electricity costs}, improving both profitability and sustainability in their operations.

\begin{figure}[H]
\centerline{\includegraphics[width=\columnwidth]{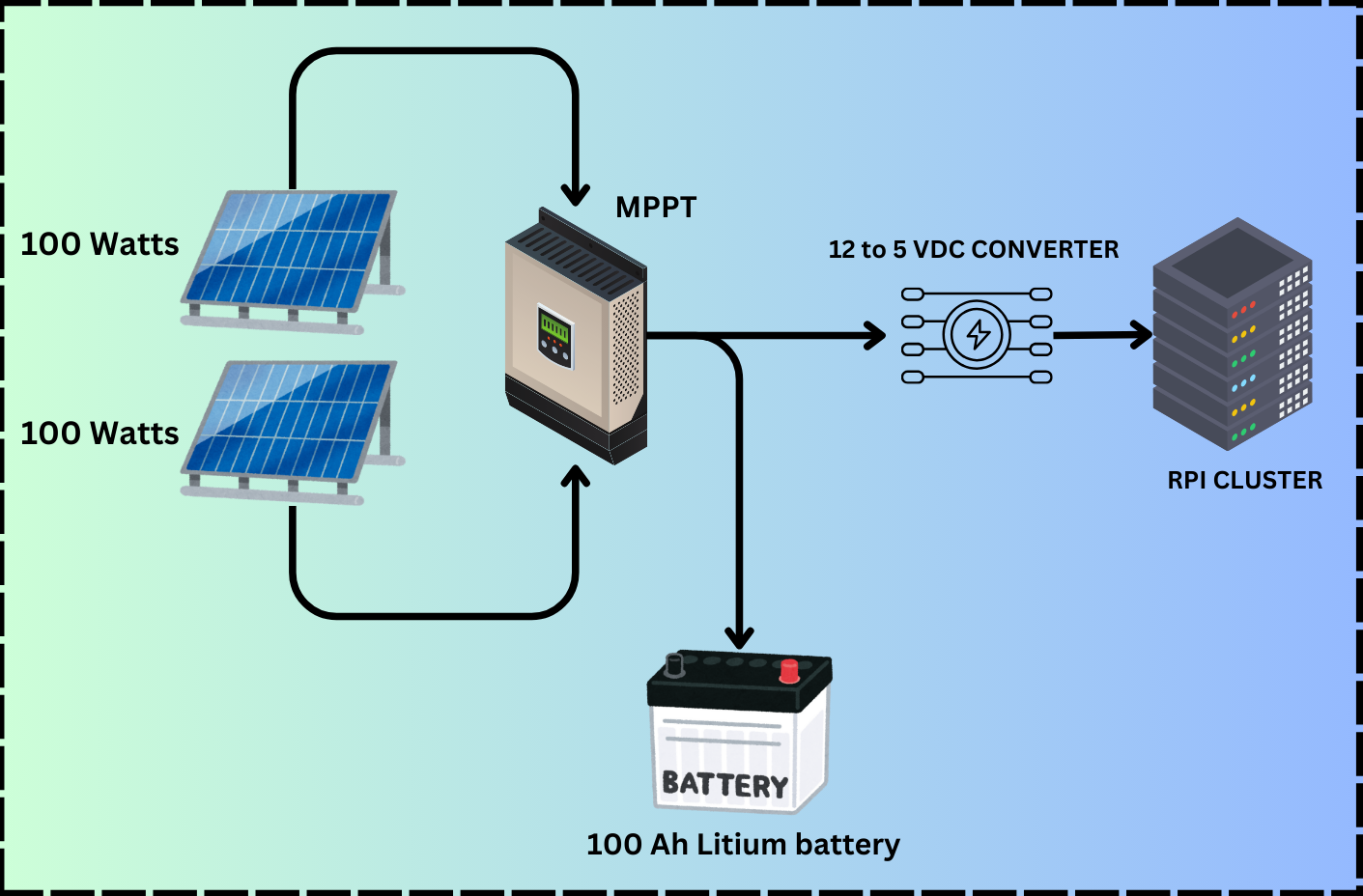}}
\captionof{figure}{Power Layer Configuration of the Prototype}
\label{prototype_power_layer_setup}
\end{figure}

\subsection{Electrical Consumption Calculations}

According to data obtained from \textit{pidramble.com} \cite{ref20}, which analyzes \textit{energy consumption benchmarks} for different versions of the \textit{Raspberry Pi board}, the \textit{Model 3B}, used in this prototype, exhibits the following power consumption levels:  

\begin{itemize}
    \item \textbf{Idle Mode:} 260 mA (1.4 W).
    \item \textbf{Moderate Load:} When executing \textit{Apache Benchmark (ab)} with \textit{100 HTTP requests and 10 concurrent connections} (\texttt{ab -n 100 -c 10}), power consumption increases to 480 mA (2.4 W).
    \item \textbf{Maximum Load:} Running \texttt{stress --cpu 4}, which subjects all four cores of the \textit{Quad-core ARM Cortex-A53 processor} to 100\% load, results in a power consumption of 730 mA (3.7 W).
\end{itemize}

\subsubsection{System Autonomy}

Under ideal solar conditions, the system's autonomy with a \textit{100Ah @ 12V battery} is summarized in Table~\ref{tab:autonomy}. However, energy efficiency losses due to DC-DC conversion (Buck converter) and battery aging must be considered, as they directly impact the actual performance of the system.
Furthermore, real-world conditions introduce variability in power availability. Cloud cover, seasonal variations, and panel degradation over time can significantly impact the system's autonomy. A deeper experimental evaluation under different environmental conditions is required to validate these estimations.

\begin{table}[h]
    \centering
    \caption{Estimated System Autonomy}
    \label{tab:autonomy}
    \renewcommand{\arraystretch}{1.3} % Mejora la legibilidad
    \begin{tabular}{|>{\raggedright\arraybackslash}m{5.2cm}|>{\centering\arraybackslash}m{2.8cm}|>{\centering\arraybackslash}m{2.8cm}|>{\centering\arraybackslash}m{2.8cm}|}
        \hline
        \multicolumn{1}{|c|}{\textbf{Load}} & \textbf{Current (A @ 5VDC)} & \textbf{Power (W @ 12V)} & \textbf{Autonomy (Hours)} \\ 
        \hline
        Idle & 1.3 A & 7.0 W & 154 h \\ 
        \hline
        Apache Benchmark (\texttt{ab -n 100 -c 10}) & 2.4 A & 12.0 W & 90 h \\ 
        \hline
        Stress CPU (\texttt{stress --cpu 4}) & 3.65 A & 18.5 W & 58 h \\ 
        \hline
    \end{tabular}
\end{table}

\textbf{Considerations:}  
\begin{itemize}
    \item The calculations are based on a cluster of \textit{5 Raspberry Pi 3 Model B units}.
    \item Autonomy is estimated considering a \textit{100Ah @ 12V battery}, including \textit{Buck converter losses} and estimated efficiency.
    \item \textbf{Conclusion:}  
    \begin{itemize}
        \item At \textit{maximum load} (CPU at 100\%),  the system is estimated to have an autonomy \textbf{2.5 days}.
        \item In \textit{idle mode}, autonomy can exceed \textbf{6 days} under optimal conditions..
    \end{itemize}
\end{itemize}

\section{Conclusions}

The research findings indicate that the \textit{IoT cluster prototype based on Raspberry Pi} and solar energy presents a promising alternative for low-cost, decentralized computing. However, further experimental validation is required to confirm its long-term feasibility in diverse environmental conditions   

The theoretical analysis of energy consumption suggests that the system can maintain its autonomy even in environments where access to the electrical grid is limited or nonexistent. The combination of a 100Ah @ 12V battery with a 200W solar panel system ensures continuous operation of the cluster under optimal conditions. However, further real-world testing is necessary to assess the long-term sustainability of the system under varying environmental conditions.

Future research should explore optimization strategies, such as the integration of ultra-low-power ARM architectures, more efficient MPPT charge controllers, advanced power management techniques, and passive cooling solutions to enhance system stability while maintaining energy efficiency.

To assess the system's sustainability, the \textit{required recharge time} to restore battery capacity was calculated based on different cluster load levels, as summarized in Table~\ref{tab:recharge_time}.  

\begin{table}[h]
    \centering
    \caption{Estimated Battery Recharge Time}
    \label{tab:recharge_time}
    \renewcommand{\arraystretch}{1.3} % Mejora la legibilidad con mayor espacio entre filas
    \resizebox{\linewidth}{!}{%
    \begin{tabular}{|>{\raggedright\arraybackslash}m{5.2cm}|>{\centering\arraybackslash}m{2.8cm}|>{\centering\arraybackslash}m{2.8cm}|>{\centering\arraybackslash}m{2.8cm}|}
        \hline
        \multicolumn{1}{|c|}{\textbf{Load}} & \textbf{Current (A @ 5VDC)} & \textbf{Power (W @ 12V)} & \textbf{Recharge Time (Hours)} \\ 
        \hline
        Idle & 1.3 A & 7.0 W & 0.196 h \\ 
        \hline
        Apache Benchmark (\texttt{ab -n 100 -c 10}) & 2.4 A & 12.0 W & 0.336 h \\ 
        \hline
        Stress CPU (\texttt{stress --cpu 4}) & 3.65 A & 18.5 W & 0.519 h \\ 
        \hline
    \end{tabular}
    }
\end{table}

These values suggest that under optimal sunlight conditions, the harvested solar energy can sustain system operation, allowing battery recharge within expected operational cycles.

The estimated recharge time assumes optimal solar conditions, with uninterrupted energy input from the photovoltaic panels. However, variations in sunlight availability, battery aging, and efficiency losses during conversion must be taken into account, f daily power demand exceeds solar energy input, additional energy storage solutions or hybrid power sources, such as grid-tied backup or secondary battery modules, may be required to maintain service continuity.
Future work should focus on testing the system under different solar exposure levels, varying workloads, and real-world deployment scenarios to assess performance and reliability over time.

\subsection{Final Considerations}
\begin{itemize}
    \item The implementation of this solution enables \textit{the decentralization of data processing}, reducing dependence on \textit{centralized infrastructures} while promoting the use of \textit{sustainable technologies}.  
    \item Its application in \textit{rural areas, off-grid locations, or mobile deployments} could represent a \textit{viable alternative to traditional cloud computing models}.  
    \item The adoption of \textit{open-source software} within the cluster architecture \textit{lowers operational costs and facilitates scalability} without reliance on proprietary licenses.  
    \item Despite its advantages, system efficiency could be further optimized through \textit{lower-power storage modules, more efficient ARM architectures, and advanced power management techniques}.  
\end{itemize}

Thus, it is concluded that the \textit{implementation of the solar-powered IoT cluster is feasible and represents a sustainable, scalable, and cost-effective alternative} for distributed computing applications in environments with limited access to traditional energy sources.

\end{document}